\documentclass[sigconf,10pt]{acmart}
\renewcommand\footnotetextcopyrightpermission[1]{} 
\usepackage{booktabs} 
\usepackage[T1]{fontenc}
\usepackage{lmodern}

\usepackage{array}
\usepackage{amssymb}
\usepackage{wrapfig}

\usepackage{graphicx}
\usepackage{url}
\usepackage{balance}

\usepackage{mathtools}

\usepackage[linesnumbered]{algorithm2e}
\usepackage{comment}

\usepackage{enumitem}

\newcommand{\eat}[1]{}
\newcommand{\Paragraph}[1]{\vspace{-0.02in}\smallskip\noindent{\bf #1.}}
\newcommand{\ParagraphNoDot}[1]{\vspace{-0.02in}\smallskip\noindent{\bf #1}}

\usepackage{fixme}
\fxsetup{
    status=draft,
    author={Note},
    layout=inline,
    theme=color,
    silent=true,
    inline=true
}
\FXRegisterAuthor{ma}{ama}{\color{red}Manos}
\FXRegisterAuthor{si}{sia}{\color{blue}Stratos}
\FXRegisterAuthor{nd}{nda}{\color{magenta}Nov}

\begin{document}
\title{Coconut Palm: Static and Streaming \\ Data Series Exploration Now in your Palm}

\author{Haridimos Kondylakis}
\affiliation{%
  \institution{\small FORTH-ICS
  }
}

\author{Niv Dayan}
\affiliation{%
  \institution{\small Harvard University
  }
}

\author{Kostas Zoumpatianos}
\affiliation{%
	\institution{\small Harvard University
	}
}

\author{Themis Palpanas}
\affiliation{%
	\institution{\small Paris Descartes University
	}
}

%
%
%


\begin{abstract}
	

Many modern applications produce massive streams of data series and maintain them in indexes to be able to explore them through nearest neighbor search. Existing data series indexes, however, are expensive to operate as they issue many random I/Os to storage.  
To address this problem, we recently proposed Coconut, a new infrastructure that organizes data series based on a new \textit{sortable} format. In this way, Coconut is able to leverage state-of-the-art indexing techniques that rely on sorting for the first time to build, maintain and query data series indexes using fast sequential I/Os.

In this demonstration, we present Coconut Palm, a new exploration tool that allows to interactively combine different indexing techniques from within the Coconut infrastructure and to thereby seamlessly explore data series from across various scientific domains. We highlight the rich indexing design choices that Coconut opens up, and we present a new recommender tool that allows users to intelligently navigate them for both static and streaming data exploration scenarios.

 

\end{abstract}

%
%



\copyrightyear{2019}
\acmYear{2019}
\setcopyright{acmcopyright}
\acmConference[SIGMOD '19]{2019 International Conference on Management of Data}{June 30-July 5, 2019}{Amsterdam, Netherlands}
\acmBooktitle{2019 International Conference on Management of Data (SIGMOD '19), June 30-July 5, 2019, Amsterdam, Netherlands}
\acmPrice{15.00}
\acmDOI{10.1145/3299869.3320233}
\acmISBN{978-1-4503-5643-5/19/06}

\maketitle

\section{Introduction}

\Paragraph{Data Series Exploration} 
Many applications today ranging from finance and multimedia to astronomy produce rapid streams of data series.
To enable domain experts to monitor and explore these series as they are created and streamed (e.g., to detect anomalies or discover patterns), it is crucial to be able to efficiently perform similarity search against user-specified query targets.

\Paragraph{Indexable Data Series Summarizations} 
Performing similarity search by comparing a target query against every individual data series 
becomes intractable as data sizes grow.
To address this problem, modern techniques compress data series into smaller summarizations 
 that allow approximating the distance to the target, and they index these summarizations such that grossly dissimilar data series can be pruned out of the search~\cite{DBLP:conf/sofsem/Palpanas16}. 
Various indexes have been proposed for this purpose including R-Tree,
$i$SAX,
ADS, 
DS-Tree, 
and SFA~\cite{DBLP:journals/pvldb/EchihabiZPB18}.
To enable interactive performance for applications, it is crucial to be able to construct, update and query such data series indexes as efficiently as possible.

\Paragraph{Problem} Existing data series indexes do not scale well.
The problem is that the summarizations on top of which these indexes are built cannot be sorted while keeping similar data series close to each other in the sorted order. 
The reason is that existing summarizations partition and tokenize data series into multiple independent segments that are laid out in their original order within the data
series.
Sorting based on these summarizations would therefore place
together data series that are similar in terms of their beginning (i.e., the first segment), yet arbitrarily far in terms of
the rest of the segments. 
Thus, state-of-the-art indexing techniques that use external sorting to create and maintain a compact and contiguous index using sequential I/Os to storage cannot be used. Instead, existing data series indexes 
are constructed using top-down insertions that lead to many random I/Os, creating a sparsely populated and non-contiguous index that requires many random I/Os to query~\cite{DBLP:journals/pvldb/KondylakisDZP18}. 

\Paragraph{Solution: Sortable Summarizations} In this work, we show how to make data series summarizations sortable. The core idea is to interleave the bits in each summarization such that the more significant bits across all segments precede all the less significant bits. As a result, sorting based on these summarizations keeps data series that are similar in terms of all of their segments close to each other in the sorted other. 

\Paragraph{Coconut Infrastructure} By making data series summarizations sortable, we built the \textbf{Co}mpact and \textbf{Con}tiguous Seq\textbf{u}ence \textbf{T}able infrastructure \cite{DBLP:journals/pvldb/KondylakisDZP18}, which leverages external sorting and log-structured updates to efficiently build and maintain a compact and contiguous data series index that is fast to query.
Coconut is extensible and can allow any state-of-the-art database indexing technique that relies on sorting to support efficient data series similarity search.  

\Paragraph{Coconut Palm} In this demonstration, we put the Coconut infrastructure at your palm. 
While there have been previous demonstrations on interactive data series exploration~\cite{Zoumpatianos2015rinse,ManninoA18}, this is the first to demonstrate the significance of compact and contiguous data layouts, thereby
further pushing the scalability envelope, not least with respect to applications with streaming data and/or with limited memory.
We provide a teaser in the URL below\footnote{\url{https://tinyurl.com/y8j35rv4}}.



\section{Coconut Indexing Versatility}

Coconut is a novel data series indexing infrastructure that organizes data series based on sortable summarizations \cite{DBLP:journals/pvldb/KondylakisDZP18}. As a result, it offers new and appealing trade-offs along various performance and space dimensions. 
In this demo, we present a new recommender tool to allow users to navigate these trade-offs and to thereby tailor an index to the specific requirements of an application: 


\Paragraph{Better Read vs. Write Trade-Offs} 
While existing data series indexes offer expensive and rigid cost balances between reads and writes on account of random I/Os, Coconut is able to harness state-of-the-art indexing techniques to achieve configurable and overall superior read/write costs by leveraging sequential I/Os to a much greater extent. 
CoconutTree (CTree), our read-optimized B-tree   implementation, 
is a compact and contiguous data series index that is extremely efficient to sequentially query, and it can further be tuned to accommodate updates by controlling its leaf nodes' fill-factor. On the other hand, CoconutLSM (CLSM), our write-optimized LSM-tree  implementation \cite{DBLP:journals/acta/ONeilCGO96}, leverages log-structured sequential writes to efficiently ingest incoming data during runtime while still providing good read performance, and it allows fine-tuning the read/write cost balance by controlling the LSM-tree's growth factor. 


\Paragraph{Better Memory vs. Construction Trade-Offs} Existing data series indexes heavily rely on in-memory buffering to alleviate the costs of index construction and maintenance by waiting for similar data series (i.e., that map to the same node) to gather, and then performing them using one I/O as a batch update. Coconut alleviates this pressure on main memory by relying on two-pass external sorting and log-structured updates to construct and maintain an index. 

\Paragraph{Better Space vs. Time Trade-Offs} As nodes in existing data series indexes are often sparsely populated, they can emerge as a storage cost bottleneck, especially on the cloud. Coconut not only alleviates such cost bottlenecks by building compact indexes, but it also provides a further option of constructing indexes as non-materialized (i.e., containing only the summarizations) or fully-materialized (i.e., also containing the original data series). 
The key trade-off  is that non-materialized indexes take up less storage and are faster to build, but subsequent queries may be slower as the raw data file has to be accessed to fetch the original data series.

\section{Coconut over Data Streams}

In data exploration scenarios, queries often have temporal constraints; they must find the nearest neighbor from within a temporal window of interest. 
In contrast to traditional streaming data applications, where values inside the window of interest are treated as sets of distinct points, we treat the values in each window as sequences of time-ordered points. 
This allows us to construct sequential patterns and query historical data as such.
We present three approaches for how to support such variable-sized window queries.

\ParagraphNoDot{Post-Processing (PP)} 
relies on examining the timestamp of every entry as it is encountered and discarding it if the timestamp falls outside the specified query window. 

\ParagraphNoDot{Temporal Partitioning (TP)} creates a new index partition based on the in-memory buffer's contents every time that the buffer fills up. 
The system gathers more and more temporal partitions over time, and it organizes them based on their creation time. 
This allows  queries to access only partitions whose creation timestamp falls within or intersects with a specified query window. 

\Paragraph{Bounded Temporal Partitioning (BTP)}  The ability to sort data series summarizations enables a new approach, Bounded Temporal Partitioning (BTP), that combines the best aspects of PP and TP. BTP creates a new temporal partition every time the buffer flushes, and it later sort-merges temporal partitions of similar sizes so that newer data resides in smaller partitions while older data gradually moves to larger contiguous partitions. As with TP, BTP allows queries over small windows to save storage bandwidth by skipping larger partitions. On the other hand as with PP, it allows queries over larger windows to spatially prune data at larger runs more effectively, and it allows approximate queries over large windows to issue fewer I/Os by bounding the overall number of partitions that need to be accessed.

\section{Coconut Palm Overview}

\begin{figure}[tb]
	\centering
	\vspace{-1mm}
	\includegraphics[width=1\columnwidth]{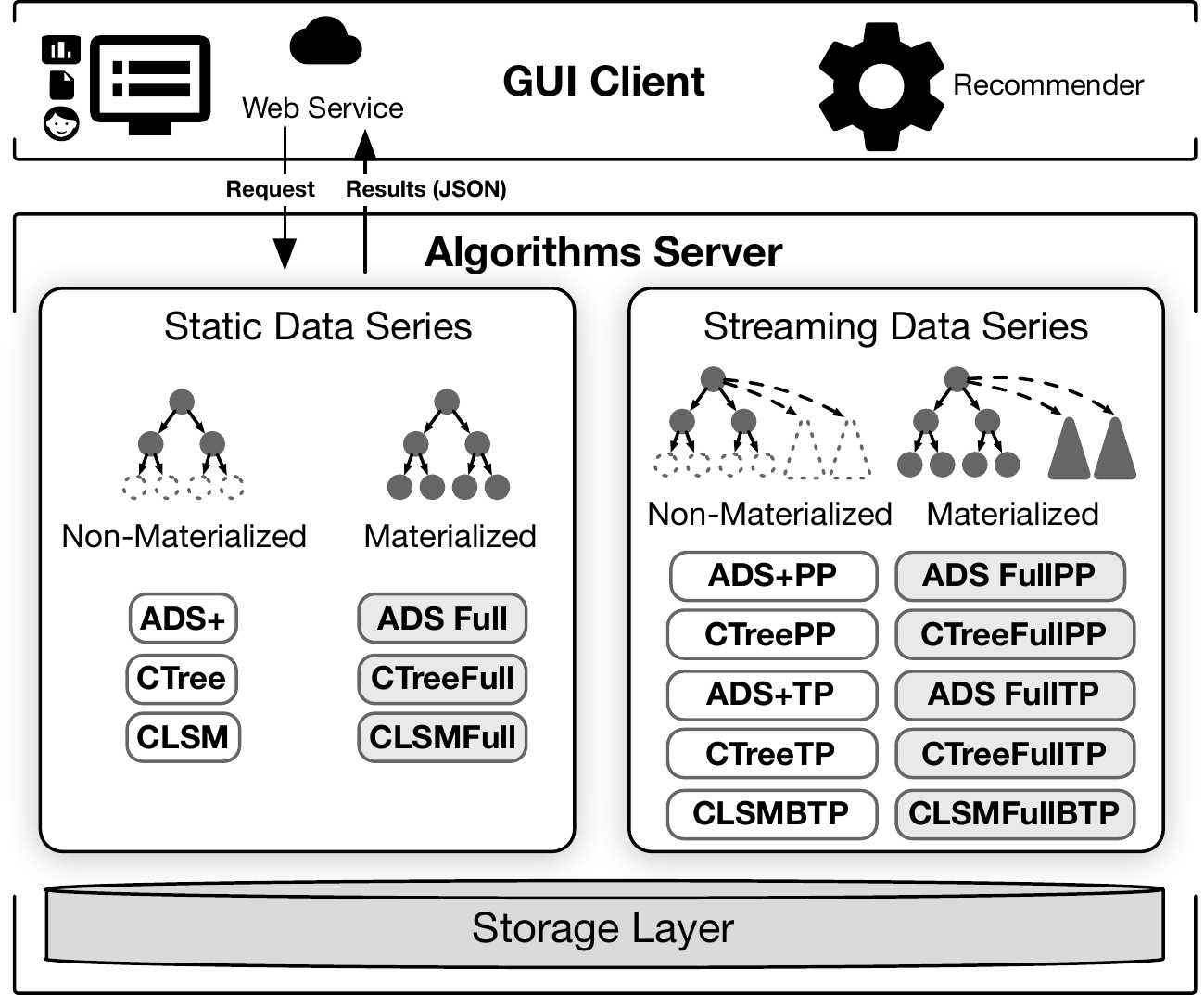}
	\vspace{-4mm}
	\caption{Coconut Palm high-level architecture.}
	\label{fig:arch}
	\vspace{-2mm}
\end{figure}

The high-level architecture of Coconut Palm is shown in Figure~\ref{fig:arch}. 
It consists of a GUI client and an algorithms server, which we describe in detail below.

\begin{figure*}[tb]
	\centering
	\vspace{-5mm}
	\includegraphics[width=1\textwidth]{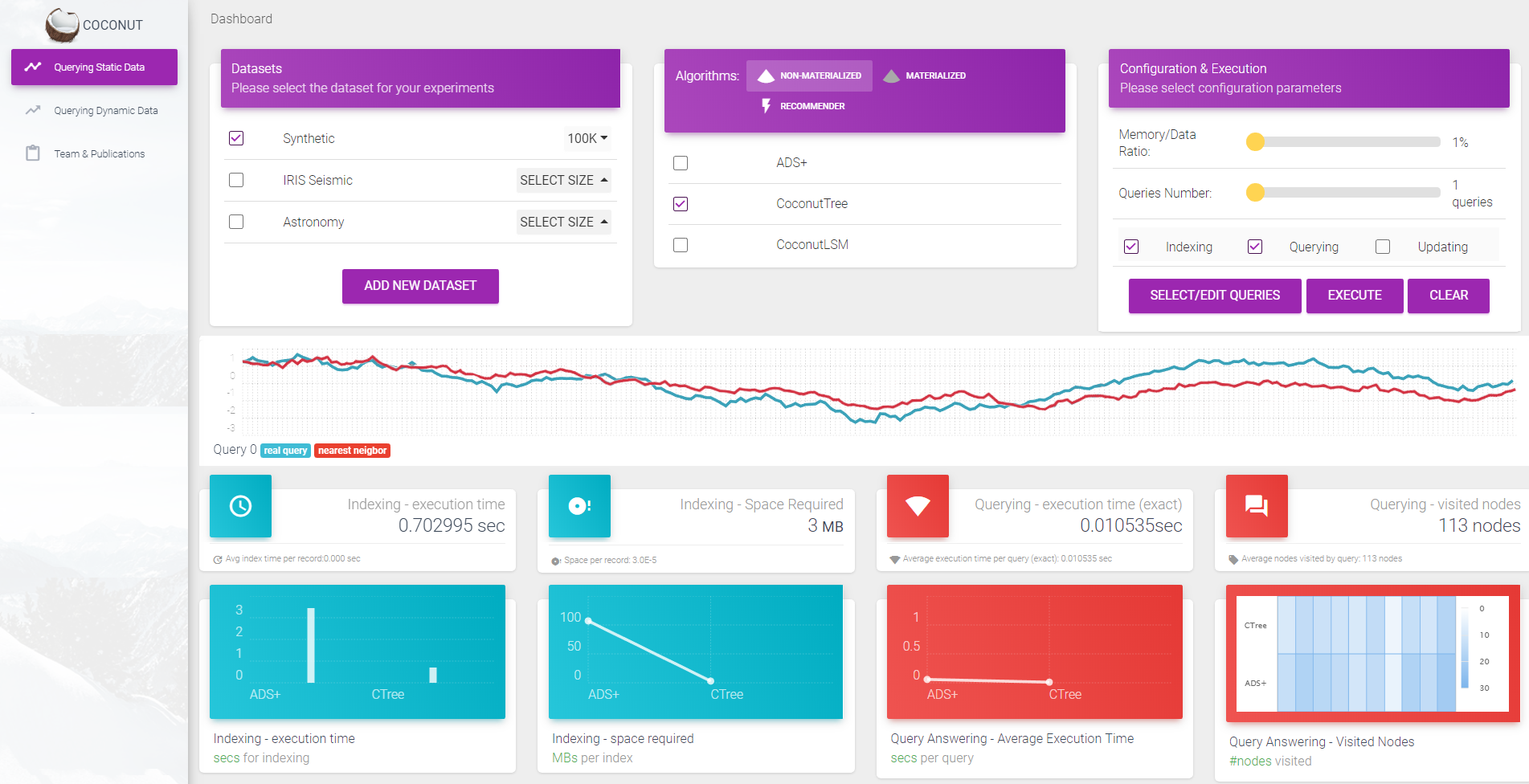}
	\vspace{-4mm}
	\caption{A screenshot of the Coconut Palm GUI.}
	\vspace{-2mm}
	\label{fig:GUI}
\end{figure*}

\Paragraph{Configurable Environment and Structures} The GUI allows users to directly interact with the Coconut infrastructure across a range of configurable application scenarios. It allows choosing between synthetic or real scientific datasets, configuring the available main memory budget, anticipating a temporal window size, and combining different algorithmic elements from the Coconut infrastructure (e.g., materialized CTree with PP) or choosing an alternative for comparison (e.g., ADSFull or ADS+). The GUI allows constructing any index of choice while visually comparing construction speed and storage consumption across index variants. A  detailed screen-shot of the GUI is shown in Figure \ref{fig:GUI}. 

\Paragraph{Recommender} Users can consult our new recommender tool for the best structural configuration for the chosen application scenario. The recommender is designed as a decision tree to be able to provide users with the rationale for its advice. 

\Paragraph{Query and Performance Visualization} Users can further draw data series, issue them as approximate or exact queries with any window size, and visually compare query performance across different index variants. To allow users to appreciate how the structural properties of an index affect query performance, we provide a heat map that visualizes a query's access pattern. 




\Paragraph{Implementation} The GUI client is developed using PHP, JavaScript and HTML. 
It communicates with a back-end server, on which the indexes are built and evaluated. 
Client-server communication takes place through REST web service calls. 
All algorithms are implemented in C/C++.

\section{Demonstration Scenarios}

There are two goals for the demonstration. The first is to show through experiments that the Coconut infrastructure significantly speeds up the process of data series exploration through faster index construction, maintenance, and querying. The second goal is to instruct users on how to choose from among the different design combinations within the Coconut infrastructure to achieve the best possible performance and space properties for a target application. We will start by guiding participants through the different design choices in Coconut. We will then walk them through two data exploration scenarios.

\Paragraph{Scenario 1: Big Static Data Series} This exploration scenario commences with a large collection of raw astronomy data series. The goal is to find data series within this dataset that match several known patterns of interest (e.g., corresponding to a supernova, a binary star, etc.). 

We will first undertake the exploration workflow with the state-of-the-art approach, ADS+, and demonstrate that it exhibits performance lags for both construction and querying. We will then consult our recommender for advice on the best Coconut index for this scenario, and we will repeat the workflow with the recommender's choice (in this case a non-materialized CTree with PP). Through first-hand experience and by visualizing performance metrics, we will demonstrate that CTree significantly speeds up the workflow. By using the heat map to analyze CTree's access patterns and comparing them to those of ADS+, we will attribute the performance improvement to more friendly I/O patterns, which are enabled as a result of constructing CTree compactly and contiguously through external sorting. 


We will show that as we increase the projected number of queries in the workload, our recommender changes its choice to using a materialized CTree, the reason being that the additional space and construction overheads for a materialized index become justified as the number of subsequent queries increases. We will allow users to issue the same set of queries to a pre-built materialized CTree on our server to appreciate the improved query performance of a materialized Coconut index. 



\Paragraph{Scenario 2: Dynamic Streaming Data Series} The second exploration scenario commences from an empty dataset and having IRIS Seismic data series\footnote{\url{http://ds.iris.edu/data/access/}} continually arrive in batches.
 The goal is to find data series that match known patterns corresponding to earthquakes from within variable-sized temporal windows of interest. 

We will use ADS+ with both PP and TP as a baseline representing the state of the art and compare it to our recommender's choice, in this case a non-materialized CLSM with BTP. By using the heat map, we will show that ongoing updates hamper the query performance of the ADS+ variants, whereas CLSM performs queries seamlessly while still being able to ingest the updates. We will further demonstrate through access pattern visualization that even in moments where updates are absent, CLSM still outperforms the ADS+ variants by virtue of using the BTP scheme to narrow the search to the partitions of interest and being able to effectively prune them to more quickly find a nearest neighbor.

\section{Conclusions}

We demonstrate Coconut, a new infrastructure that accelerate the process of data series exploration. 
The core innovation is a \textit{sortable} data series summarization, which allows using state-of-the-art indexing techniques for the first time 
to efficiently construct, maintain and query a data series index. 
We demonstrate the versatile new performance and space trade-offs that Coconut provides,
and we allow users to experience and navigate the infrastructure with the aid of a new recommender tool. 




\section{Acknowledgments}

This work was partially supported by the projects BOUNCE (H2020 \#777167), myPal (H2020 \#825872), NESTOR (Marie Curie \#748945), and the FMJH Program PGMO (in cooperation with EDF-THALES).

\bibliographystyle{abbrv}
\bibliography{sample-bibliography}

\begin{thebibliography}{1}

\bibitem{DBLP:journals/pvldb/EchihabiZPB18}
K.~Echihabi, K.~Zoumpatianos, T.~Palpanas, and H.~Benbrahim.
\newblock The lernaean hydra of data series similarity search: An experimental
  evaluation of the state of the art.
\newblock {\em {PVLDB}}, 12(2), 2018.

\bibitem{DBLP:journals/pvldb/KondylakisDZP18}
H.~Kondylakis, N.~Dayan, K.~Zoumpatianos, and T.~Palpanas.
\newblock Coconut: {A} scalable bottom-up approach for building data series
  indexes.
\newblock {\em {PVLDB}}, 11(6):677--690, 2018.

\bibitem{ManninoA18}
M.~Mannino and A.~Abouzied.
\newblock Qetch: Time series querying with expressive sketches.
\newblock In {\em SIGMOD}, 2018.

\bibitem{DBLP:journals/acta/ONeilCGO96}
P.~E. O'Neil, E.~Cheng, D.~Gawlick, and E.~J. O'Neil.
\newblock The log-structured merge-tree (lsm-tree).
\newblock {\em Acta Inf.}, 33(4), 1996.

\bibitem{DBLP:conf/sofsem/Palpanas16}
T.~Palpanas.
\newblock Big sequence management: {A} glimpse of the past, the present, and
  the future.
\newblock In {\em SOFSEM}, 2016.

\bibitem{Zoumpatianos2015rinse}
K.~Zoumpatianos, S.~Idreos, and T.~Palpanas.
\newblock {RINSE:} interactive data series exploration with {ADS+}.
\newblock {\em {PVLDB}}, 8(12), 2015.

\end{thebibliography}

\end{document}